\documentclass[iop]{emulateapj}

\received{2010 December 9}
\accepted{2011 January 25}
\usepackage{epsfig}

\def\simgt{\lower.5ex\hbox{$\; \buildrel > \over \sim \;$}}
\def\simlt{\lower.5ex\hbox{$\; \buildrel < \over \sim \;$}}

\def\etal{{et~al.}}
\def\amin{\ifmmode^{\prime}\else$^{\prime}$\fi}
\def\asec{\ifmmode^{\prime\prime}\else$^{\prime\prime}$\fi}

\def\simgt{\lower.5ex\hbox{$\; \buildrel > \over \sim \;$}}
\def\simlt{\lower.5ex\hbox{$\; \buildrel < \over \sim \;$}}

\newcommand\xte{{\it RXTE\/}}

\newcommand\chandra{{\it Chandra}}

\newcommand\xmm{{\it XMM-Newton}}

\newcommand\integral{{\it INTEGRAL}/IBIS}

\newcommand\swift{{\it Swift\/}}

\def\startd{2010 November~25}
\def\endd{December~15}
\def\rxtetime{112.3~ks}
\def\epoch{55535.285052933}
\def\mjdrangev{55525.6 -- 55545.2}
\def\periodv{38.518943432(12)~ms}
\def\pdotv{$1.4224(58) \times 10^{-14}$ s~s$^{-1}$}
\def\edotv{$9.8 \times 10^{36}$ erg~s$^{-1}$}
\def\bsv{$7.5\times 10^{11}$~G}
\def\taucv{42.9~kyr}

\def\period{38.5~ms}

\def\pdotl{$1.42 \times 10^{-14}$ s~s$^{-1}$}
\def\edot{$\dot E =$ \edotv}
\def\bs{$B_s =$ \bsv}
\def\tauc{$\tau_c \equiv P/2\dot P =$ \taucv}

\def\tev{HESS~J1849$-$000}
\def\xmmu{XMMUU~J184901.6$-$000117}

\def\igr{IGR~J18490$-$0000}

\def\xmmu{XMMU~J184901.6$-$000117}
\newcommand\psr{PSR~J1849$-$0001}

\slugcomment{To Appear in The Astrophysical Journal Letters}

\shorttitle{Discovery of a Pulsar Powering \igr}
\shortauthors{Gotthelf et al.}

\begin{document}

\title{Discovery of an Energetic \period\ Pulsar Powering
the Gamma-ray Source \igr/\tev}

%Discovery of an Energetic \period\ Pulsar Powering the Gamma-ray Source IGR J18490-0000/HESS J1849-000}

\author{E. V. Gotthelf\altaffilmark{1}, J. P. Halpern\altaffilmark{1},
R. Terrier\altaffilmark{2} and F. Mattana\altaffilmark{2}}

\altaffiltext{1}{Columbia Astrophysics Laboratory, Columbia University,
550 West 120th Street, New York, NY 10027, USA; eric@astro.columbia.edu}
\altaffiltext{2}{Astroparticule et Cosmologie, Universit\'e Paris 7/CNRS/CEA,
Batiment Condorcet, 75013 Paris, France}

%We report the discovery of a 38.5 ms X-ray pulsar in observations of
%the soft gamma-ray source IGR J18490-0000 with the Rossi X-ray Timing
%Explorer (RXTE). PSR J1849-0001 is spinning down rapidly with period
%derivative 1.42E-14 s/s, yielding a spin-down luminosity 9.8E36
%erg/s, characteristic age 42.9 kyr, and surface dipole magnetic field
%strength 7.5E11 G. Within the INTEGRAL/IBIS error circle lies a
%point-like XMM-Newton and Chandra X-ray source that shows evidence of
%faint extended emission consistent with a pulsar wind nebula (PWN).
%The XMM-Newton spectrum of the point source is well fitted by an
%absorbed power-law model with photon index Gamma(PSR) = 1.1 +/- 0.2,
%N_H = (4.3+/-0.6)E22 cm^-2, and F(PSR;2-10keV) = (3.8+/-0.3)E-12
%erg/s/cm^2, while the spectral parameters of the extended emission are
%Gamma(PWN) = 2.1 and F(PWN;2-10 keV) = 9E-13 erg/s/cm^2.  IGR
%J18490-0000 is also coincident with the compact TeV source HESS
%J1849-000. For an assumed distance of 7 kpc in the Scutum arm tangent
%region, the 0.35-10 TeV luminosity of HESS J1849-000 is 0.13% of the
%pulsar's spin down energy, while the ratio F(0.35-10 TeV)/F(PWN; 2-10
%keV) of approx. 2.  These properties are consistent with leptonic
%models of TeV emission from PWNe, with PSR J1849-0001 in a stage of
%transition from a synchrotron X-ray source to an inverse Compton
%gamma-ray source.

\begin{abstract}

We report the discovery of a \period\ X-ray pulsar in observations of
the soft $\gamma$-ray source \igr\ with the {\it Rossi X-ray Timing
Explorer} (\xte).  \psr\ is spinning down rapidly with period
derivative \pdotl, yielding a spin-down luminosity
\edot, characteristic age \tauc, and surface dipole magnetic field 
strength \bs. Within the \integral\ error circle lies a point-like
\xmm\ and \chandra\ X-ray source that shows evidence of faint extended emission 
consistent with a pulsar wind nebula (PWN).  The \xmm\ spectrum of the point source is
well fitted by an absorbed power-law model with photon index
$\Gamma_{PSR} = 1.1 \pm 0.2$, $N_{\rm H} = (4.3 \pm 0.6)
\times 10^{22}$~cm$^{-2}$, and $F_{\rm PSR}(2-10\,{\rm keV}) = (3.8 \pm 0.3) \times
10^{-12}$ erg~cm$^{-2}$~s$^{-1}$, while the spectral parameters of the
extended emission are roughly $\Gamma_{\rm PWN} \approx 2.1$ and $F_{\rm PWN}(2-10\,{\rm
keV}) \approx 9 \times 10^{-13}$ erg~cm$^{-2}$~s$^{-1}$.  
\igr\ is also coincident with the compact TeV source \tev. For an
assumed distance of 7~kpc in the Scutum arm tangent region, the
$0.35-10$~TeV luminosity of \tev\ is 0.13\% of the pulsar's spin-down energy,
while the ratio $F(0.35-10\,{\rm TeV})/F_{\rm PWN}(2-10\,{\rm keV})
\approx 2$.  These properties are consistent with leptonic models of
TeV emission from PWNe, with \psr\ in a stage of transition from a
synchrotron X-ray source to an inverse Compton $\gamma$-ray source.
\end{abstract}

\section{Introduction}

The detection of $10^{12}$~eV emission associated with supernova
products in the Galaxy has opened up a new window on the evolution of
these energetic stellar remnants.  More than 2/3 of the now $>60$
Galactic TeV sources are supernova remnants or pulsar wind nebulae
(PWNe), the latter being the largest class\footnote{VHE $\gamma$-ray
Sky Map and Source Catalog,\\
\url{http://www.mppmu.mpg.de/$\sim$rwagner/sources/index.html}}.  Many
of these TeV PWNe are spatially offset from middle-aged 
($\tau_c \sim 10^{4}-10^{5}$~year old) pulsars,
and are often more extended and more
luminous than their X-ray PWNe counterparts. 
This TeV emission is likely to be inverse Compton scattered radiation
from relic electrons produced by the pulsar in an earlier stage
of energetic spin-down \citep{dej08,zha08,mat09b}
In contrast, younger ($\tau_c \sim 10^3$~year old)
pulsars are associated with compact
TeV sources that are co-located with their X-ray PWNe. 
In these younger systems the high magnetic fields make for efficient
synchrotron X-ray sources, and inefficient inverse Compton TeV emission.

\tev\ is one of the fainter sources detected in the HESS Galactic
Plane survey \citep{aha05,aha06} and subsequent dedicated observations,
with significance of $6.4\sigma$, flux $>350$~GeV of $\approx
15$~mCrab, and only weak indication of extent \citep{ter08}.
\tev\ is coincident in position with the \integral\ source \igr\
that was discovered during a survey of the tangent regions of
the Sagittarius and Scutum spiral arms
\citep{mol04,bir06}. Follow-up X-ray observation of
\igr\ by \citet{rod08} with the \swift\ X-ray telescope
(XRT) located a highly absorbed X-ray point source within the
\integral\ error circle.  These authors suggested a Galactic X-ray
binary origin for \igr\ based on a $K$-band Two Micro All Sky Survey
star located within the \swift\ XRT error circle. However, that
association was excluded by the precise X-ray position obtained in a
brief \chandra\ High Resolution Camera (HRC) observation
\citep{rat10}.
\citet{ter08} used imaging and spectroscopic evidence from an
\xmm\ observation to argue that \igr\ is instead a pulsar/PWN.
As such, it would join the dozen hard X-ray members of this
class detected by \integral\ \citep{mat09a,ren10}.

In Section~2 we present in detail the \xmm\ imaging and spectral data
that support a pulsar/PWN interpretation for \igr.  In Section~3, we
report the discovery using {\it Rossi X-ray Timming Explorer} (\xte)
of \psr\ and its spin-down, which verifies this conjecture.  In
Section~4, we discuss the properties of \tev\ in the context of the
spin-down parameters of \psr.

\section{\xmm\ Observation}

\begin{figure*}[t]
\centerline{
%\psfig{figure=igr1849_hrc.ps,height=0.32\linewidth,angle=270}
%\hfill
\psfig{figure=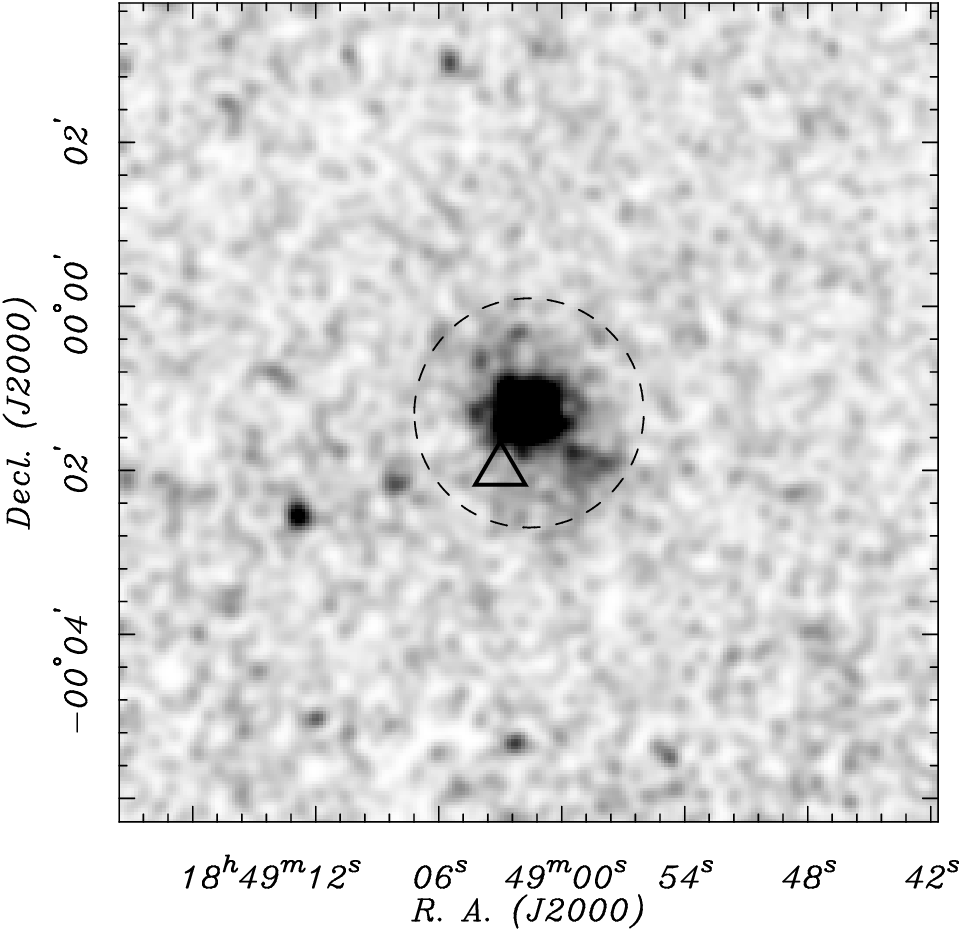,height=0.48\linewidth,angle=270}
\hfill
\psfig{figure=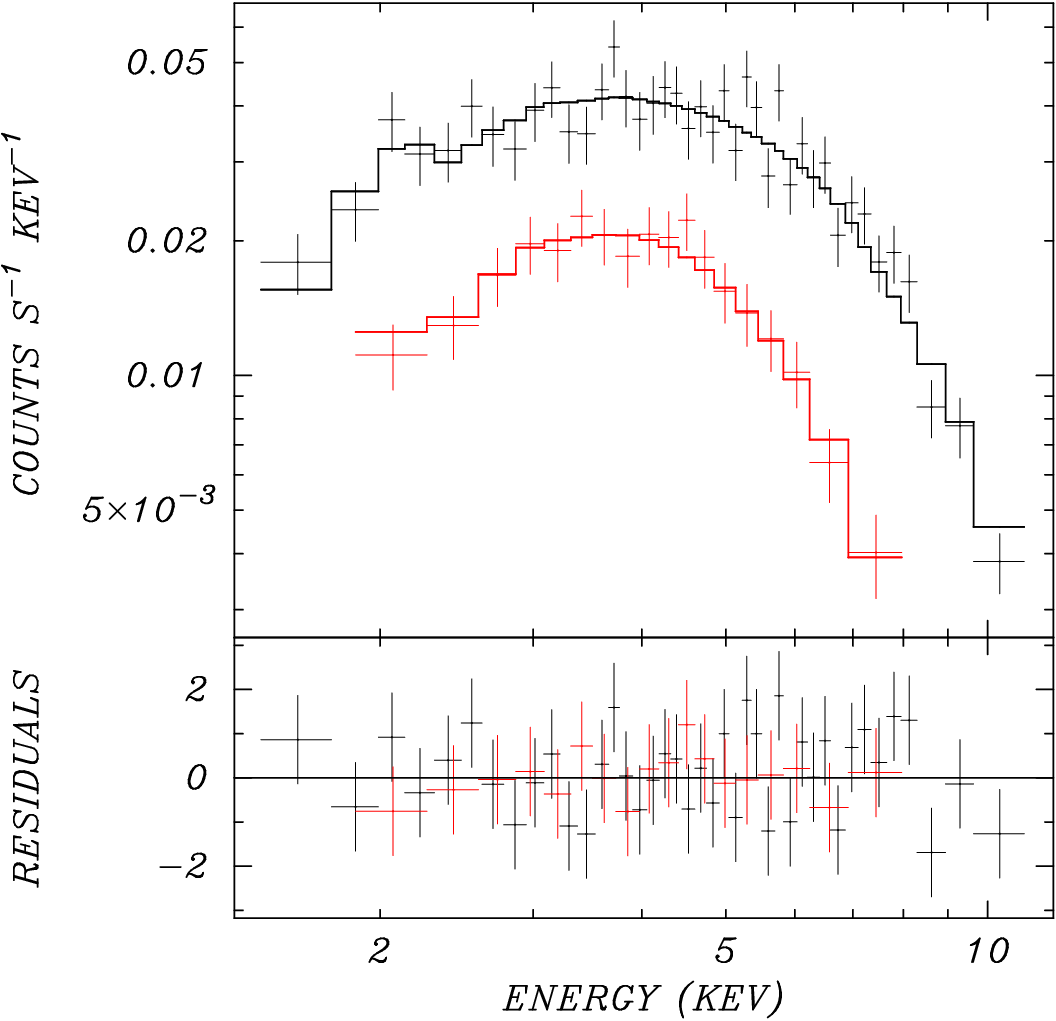,height=0.48\linewidth,angle=270}
}
\caption{
\small \xmm\ EPIC X-ray observation of the \integral\ source \igr.
Left: exposure corrected and smoothed EPIC MOS image scaled in
intensity to highlight the faint diffuse emission near the point
source. This source, the putative pulsar, lies within the \integral\
$r=1\farcm4$ positional uncertainty of \igr\ (\citealt{ter08}; dashed
circle), as does the centroid of \tev, indicated by the triangle.
Right: \xmm\ EPIC~pn (black) and EPIC~MOS (red) spectra of the
point source fitted to the model described in the text. The lower
panel shows residuals from the best fit in units of $1\sigma$.}
\label{fig:images}
\end{figure*}

An 11~ks \xmm\ observation of \igr\ (ObsID 0306170201) was acquired on
2006 April 3 using the European Photon Imaging Camera (EPIC;
\citealt{tur03}). EPIC consists of three sensors operating in
parallel, one pn and two MOS CCD cameras. These instruments are
sensitive to X-rays in the 0.2--12~keV range with energy resolution
$\Delta E/E \approx 0.1/\sqrt{E({\rm keV})}$.  All three cameras were
operated in full-frame mode, using the thin and medium filters for the
EPIC pn and MOS, respectively.  The target was placed at the default
EPIC~pn focal plane location for a point source.  The time resolution
of 73.4~ms and 2.7~s for the EPIC pn and MOS, respectively, are
insufficient to search for a typical pulsar signal.

The data were processed using the SAS version
xmmsas\_20061026\_1802-6.6.0 pipeline, and were analyzed using both the
SAS and FTOOLS software packages. The observation was free of
significant particle background contamination and provided a near
continuous 11~ks of good observing time for the EPIC~MOS and 9.9~ks
for EPIC~pn.

Figure~\ref{fig:images} (left) displays the 2--10~keV \xmm\ EPIC~MOS
X-ray image centered on \igr. The image has been exposure corrected
and smoothed using a $\sigma = 3\farcs8$ Gaussian kernel and scaled to
highlight the faint diffuse emission near the central bright source at
coordinates (J2000.0) R.A. = $18^{\rm h}49^{\rm m}01.\!^{\rm s}62$,
decl. = $-00^{\circ}01^{\prime}17.\!^{\prime\prime}7$.  This source,
at the position of the previously detected \swift\ XRT source, lies
at the center of the 1\farcm4 radius error circle of \igr. There are
no other prominent X-ray sources in the full $30^{\prime}$ diameter
field of this instrument. Together with the positional coincidence, its
flux and spectrum (see below) leave no doubt that \xmmu\ is the
counterpart of \igr.  Faint, diffuse emission surrounds the point
source, and a prominent feature extends up to $2\!^{\prime}$ to the
southwest; these properties are suggestive of a PWN.  A
1.2~ks \chandra\ HRC image obtained on 2008 February 16 \citep{rat10}
shows that most of the \xmm\ source flux is contained in a point-like
component.

\subsection{\xmm\ Point-source Spectrum}

A spectrum of \xmmu\ was extracted from each EPIC camera using a
$30^{\prime\prime}$ radius aperture centered on the point source; data
from the two MOS cameras were combined into a single spectrum. The
source background was estimated using counts from an
$60^{\prime\prime}$ aperture placed just south of the source region,
on the same CCD of each camera. These source spectra were grouped with
a minimum of 50 counts per spectral channel and fitted using the XSPEC
fitting package \citep{arn96}. EPIC pn and MOS spectra were fitted
simultaneously with a single absorbed power-law model, allowing
independent flux normalization (Figure~\ref{fig:images}, right); we
report the average of the fluxes measured by the two EPIC cameras. In
the following, all derived luminosities are corrected for interstellar
absorption, fluxes are uncorrected, and errors are at the 90\%
confidence level. The data yield a good fit with a reduced
$\chi^2_{\nu}=0.90$ for 72 degrees of freedom (dof). The hard spectrum
has photon index $\Gamma_{\rm PSR} = 1.1 \pm 0.2$, $N_{\rm H} = (4.3 \pm
0.6) \times 10^{22}$~cm$^{-2}$, and $F_{\rm PSR}(2-10\, {\rm keV}) =
(3.8\pm0.3)
\times 10^{-12}$ erg~cm$^{-2}$~s$^{-1}$.  The Galactic coordinates of
\igr, $(\ell,b) = (32.\!^{\circ}64,+0.\!^{\circ}53)$ are coincident
with the tangent point of the Scutum arm.  Considering this and the
large fitted $N_{\rm H}$, we hypothesize that \igr\ is located at a
distance of 7~kpc appropriate for the Scutum tangent region rather
than residing in the nearer Sagittarius arm.  Its $2-10$~keV X-ray
luminosity is then $L_{\rm PSR}(2-10\,{\rm keV})= 2.8
\times 10^{34}\,d_{7}^2$~erg~s$^{-1}$.
 
We also fitted the \xmm\ and \integral\ spectra simultaneously,
finding that a simple power-law model gives a reasonable $\chi^2$ but
a rather poor fit to the \integral\ data (observation and analysis of
the \igr\ data as described in Terrier \etal\ 2008). An excellent
result is obtained by fitting a broken power law ($\chi^2_{\nu}=0.95$ for
140 dof) with a break energy of $23$~keV. The best fit indices of
$\Gamma= 1.2\pm 0.1$ and $\Gamma = 1.7\pm 0.2$, for the low and high
energies, respectively, are consistent with those obtained for the
spectra individually \citep{ter08}.
%suggests that there is spectral break at $\approx 20$ keV, or a gradual
%steepening of the spectrum 

\subsection{\xmm\ Extended Emission}

\begin{figure}
\vspace{-0.15in}
\includegraphics[height=0.36\textheight]{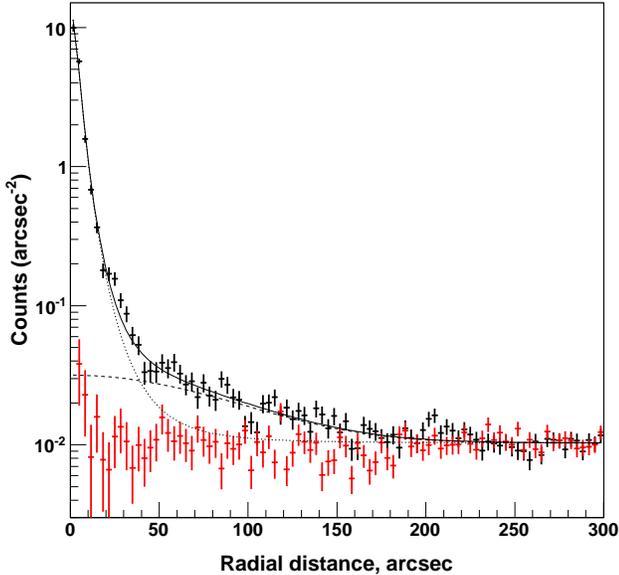}
\caption{Radial profile of the emission from \igr\ in the EPIC pn.
The dotted line is the average King profile for a $\Gamma=1.1$
power-law spectrum; the red points are background taken at a symmetric
position with respect to the instrument optical axis.  An additional
extended component visible up to
$150^{\prime\prime}$ from the point source is
well fitted by a Gaussian profile of $\sigma=75^{\prime\prime}$
(dashed line).  The solid line is the total fitted profile.}
\label{fig:profile}
\end{figure}

To quantify the spatial structure of the \xmm\ source,
we made a radial profile of the counts
distribution centered on the point source and compared it with the point
spread function (PSF).  The latter was constructed using
instrumental profiles at several
energies weighted to match the measured spectral index.
Figure~\ref{fig:profile} shows the measured EPIC pn profile compared
to the average PSF for this spectrum and location in the field.
Background has been taken from a symmetric position with respect
to the optical axis.  There is a clear excess of emission from
$20^{\prime\prime}$ to $150^{\prime\prime}$ radius.  We fitted the radial
counts distribution to the sum of a King profile to represent the
point source, and a Gaussian to represent the extended emission.
A Gaussian component $150^{\prime\prime}\pm 15^{\prime\prime}$
in diameter improves the quality of the fit by $\approx 9\sigma$,
proving that the source is surrounded by a nebula or a halo.

We extracted the spectrum of the extended nebula in an annulus
surrounding \xmmu\ from $30^{\prime\prime}$ to $150^{\prime\prime}$.
To account for contamination from the wings of the bright point
source, we fitted the point source and extended emission spectra
simultaneously.  We used separate power laws but the same $N_{\rm H}$
for both spectra.  We added a constant fraction of the point-source
spectrum to the diffuse emission. This fraction was estimated from the
PSF as 12\% and 14\% for EPIC pn and MOS, respectively, at a radius of
$30^{\prime\prime}$.  To account for the uncertainty in the estimate
of these fractions, we varied them from 7\% to 19\% of the
point-source flux, adding the corresponding systematic uncertainty to
errors on the fitted parameters. The resulting diffuse emission
spectrum has $\Gamma_{\rm PWN} = 2.1 \pm 0.3$ and a flux of
$F_{\rm PWN}(2-10\,{\rm keV}) = (9 \pm 2)\times 10^{-13}$
erg~cm$^{-2}$~s$^{-1}$.  The reduced chi-squared is $\chi^2_{\nu} =
0.90$ for 311 dof.

The extended nebula with a steeper spectrum suggests that \igr\ is a
pulsar/PWN system.  Alternatively, given the large $N_{\rm H}$, some
of the diffuse emission could be a dust-scattered halo of the bright
point source \citep{pre95}; however, that could not account for its
asymmetric extension.  We note that the nebula is rather faint,
$\approx 25\%$ of the point source flux, but it is still in the range
observed for PWNe \citep{kar08}.

\section{\xte\ Observations}

To search for the expected pulsar signal from \igr, we obtained
\rxtetime\ of \xte\ observations, pointed at the \chandra\ source position,
spanning \startd\ -- \endd.
%A log of the
%observations is given in Table~\ref{tab:xtelog}.
The schedule was designed to obtain a phase-connected timing solution
that would measure both $P$ and $\dot P$.  The data used here
were collected with the Proportional Counter Array
\citep[PCA;][]{jah96} in the GoodXenon mode with an average of 1.75 out
of the five proportional counter units (PCUs) active. In this mode,
photons are time-tagged to $0.9$ $\mu$s precision and have an absolute
time accuracy of $<100\,\mu$s \citep{rot98}. The effective area
of five combined detectors is $\approx 6500$ cm$^{2}$ at 10~keV with a
roughly circular field of view of $\sim 1^{\circ}$ FWHM. Spectral
information is available in the 2$-$60~keV energy band with a
resolution of $\sim 16\%$ at 6~keV.

Standard time filtering was applied to the PCA production data,
rejecting intervals of South Atlantic Anomaly passage, Earth
occultation, and other periods of high particle activity.  The photon
arrival times were transformed to the solar system barycenter in
Barycentric Dynamical Time (TDB) using the JPL DE200 ephemeris and the
coordinates given in Table~\ref{tab:ephem}, obtained from \chandra\
HRC observation ObsID~7398. These coordinates are derived from a
centroid calculation of 17~photons within a $2^{\prime\prime}$ radius
aperture and have nominal uncertainty of 0\farcs6. We note that these
coordinates differ by $0\farcs35$ from those reported in
Ratti \etal\ (2010).

\subsection{\xte\ Timing Analysis}

\begin{deluxetable}{ll}
%\tabletypesize{\small}
\tablecaption{Timing Parameters of \psr }
%\tablecolumns{2}
%\tablewidth{0pc}
\tablehead{
\colhead{Parameter}   &
\colhead{Value}
}
\startdata                                       
R.A. (J2000.0)\tablenotemark{a} \dotfill    & $18^{\rm h}49^{\rm m}01.\!^{\rm s}61$ \\
Decl. (J2000.0)\tablenotemark{a} \dotfill   & $-00^{\circ}01^{\prime}17.\!^{\prime\prime}6$ \\
Epoch (MJD TDB) \dotfill                    & \epoch\  \\
Period\tablenotemark{b}, $P$ \dotfill                 & \periodv\  \\
Period derivative\tablenotemark{b}, $\dot P$ \dotfill & \pdotv\ \\
Range of dates (MJD) \dotfill                 & \mjdrangev\   \\
Spin-down luminosity, $\dot E$ \dotfill       & \edotv\           \\
Characteristic age, $\tau_c$ \dotfill         & \taucv\           \\
Surface dipole magnetic field, $B_s$ \dotfill & \bsv
\enddata
%\tablenotetext{a}{\footnotesize Chandra HRC position from \citet{rat10}.}
\tablenotetext{a}{\footnotesize \chandra\ position with a nominal uncertainty of 0\farcs6 (see the text).}
\tablenotetext{b}{\footnotesize TEMPO $1\sigma$ uncertainties given
in parentheses.}
\label{tab:ephem}
\end{deluxetable}

\begin{figure*}[t]
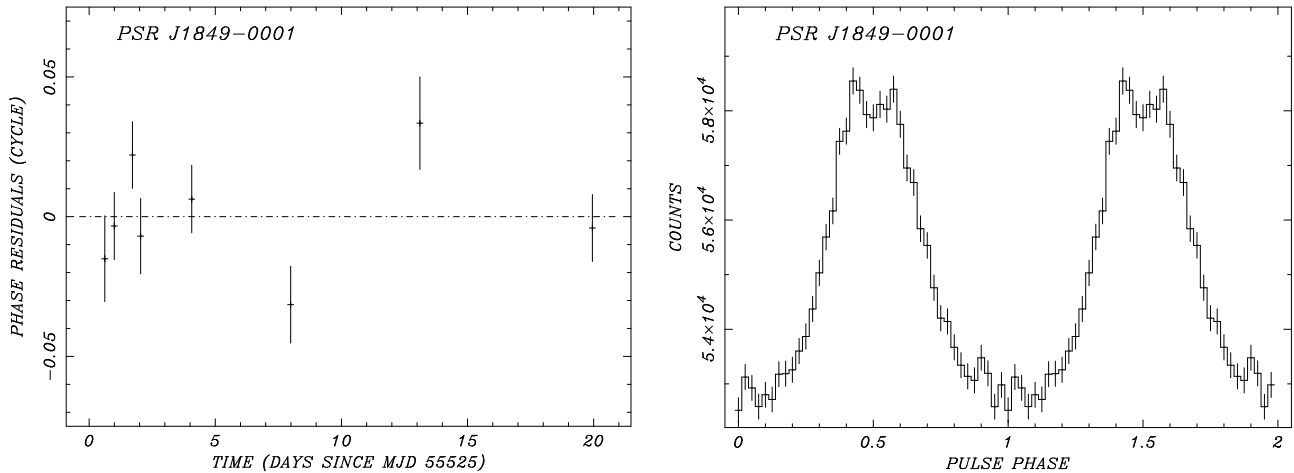

\centerline{
\hfill
\includegraphics[height=0.46\linewidth,angle=270,clip=true]{f3.eps}
\hfill
\includegraphics[height=0.46\linewidth,angle=270,clip=true]{f4.eps}
\hfill
}
\caption{Discovery timing results for the X-ray pulsar \psr\ using \xte\ data.
Pulse phase residuals { (left)} and folded light curve {(right)} of
\psr\ in the 2$-$20 keV band were generated using the quadratic
ephemeris of Table~\ref{tab:ephem}.  Phase zero of the light curve
corresponds to the ephemeris epoch; two cycles are shown for clarity.}
\label{fig:xtepulse}
\end{figure*}

We restricted the analysis to the $2-20$~keV energy range (PCA
channels $2-50$) from the top Xenon layer of each PCU to optimize the
signal-to-noise.  A fast Fourier transform of the initial data set
(ObsID 95309-01-01-02; 7~ks) revealed a highly significant signal of
period $P=38.52$~ms.  Subsequently, for each satellite orbit we
extracted a pulse profile corresponding to the peak power as
determined by the Rayleigh test \citep{str80}, also known as $Z^2_1$
\citep{buc83}. The resulting profiles were cross-correlated, shifted,
and summed to generate a master pulse profile template.  Each
individual profile was then cross correlated with this template to
determine its time of arrival (TOA) and uncertainty.  These TOAs were
then iteratively fitted to a quadratic ephemeris.

The resulting unique ephemeris is presented in Table~\ref{tab:ephem},
and the phase residuals are shown in Figure~\ref{fig:xtepulse}.
The residuals are all less than 0.05 cycles and do not
require a higher derivative.
The derived properties of \psr\ are:
spin-down luminosity $\dot E = 4\pi^2I\dot P/P^3 =$ \edotv,
characteristic age $\tau_c \equiv P/2\dot P =$ \taucv,
and surface dipole magnetic field strength
$B_s = 3.2 \times 10^{19}(P\dot P)^{1/2} =$ \bsv.

Figure~\ref{fig:xtepulse} displays the summed pulse profile using all
the 2$-$20 keV data folded on the final ephemeris.  It has a
symmetric, single-peaked structure.  The raw pulsed fraction is
$3.8\%$, uncorrected for PCA instrument or astrophysical background,
difficult to determine with sufficient accuracy for this purposes.
We see no energy dependence of the pulse profile when subdividing the
2--20~keV band.

\subsection{\xte\ Spectral Analysis}

The spectrum of the pulsed flux from \psr\ can be isolated by
phase-resolved spectroscopy. We used the {\it fasebin} software to
construct phase-dependent spectra based on the ephemeris of
Table~\ref{tab:ephem}.  For each ObsID and PCU we constructed spectra
from counts detected in the top Xenon layer only and combined them to
produce a single spectrum per PCU for the entire set of observations.
Similarly, standard PCA responses for each PCU were generated at each
ObsID and averaged.  In fitting the pulsed flux, the unpulsed emission
provides a near perfect background estimate and was taken from the 0.4
phase bins range corresponding to the region of flat minimum in the
pulse profile around phase zero.

The merged \xte\ spectrum was fitted in the $2-20$~keV range using a
simple absorbed power-law model with the interstellar absorption held
fixed at $N_{\rm H} = 4.3\times10^{22}$~cm$^{-2}$ determined from the
\xmm\ fit (see Figure 4).  The resulting best-fit photon index is $\Gamma =
1.3^{+0.2}_{-0.3}$. The 2$-$10 keV pulsed flux is $8.9 \times
10^{-13}$~erg~cm$^{-2}$~s$^{-1}$, which represents $\sim 25\%$ of the
point-source flux of \psr\ measured by \xmm.  The corresponding pulsed
luminosity (assumed isotropic) is $6.7 \times
10^{33}\,d_{7}^2$~erg~s$^{-1}$.  The reduced chi-squared is
$\chi^2_{\nu} = 0.70$ for 29 dof.

As expected, the \xte\ measured spectral slope is intermediate
between the \xmm\ and \integral\ values, indicating a steepening
around $10-20$~keV.

%\begin{figure}[t]
%\centerline{
%\includegraphics[height=0.96\linewidth,angle=270,clip=true]{f4.eps}}
%\caption{\xte\ folded light curve of \psr\ in the 2$-$20 keV band.
%Phase zero corresponds to the epoch of the ephemeris in Table~\ref{tab:ephem}.}
%\label{fig:xtepulse}
%\end{figure}

\begin{figure}[t]
\centerline{
\includegraphics[height=0.97\linewidth,angle=270,clip=true]{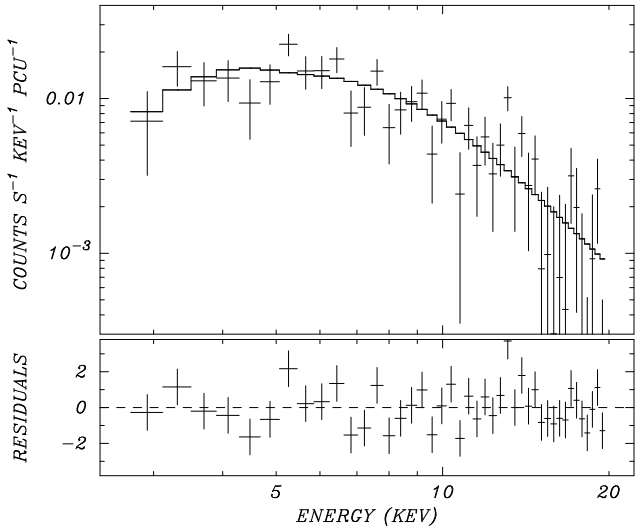}
}
%\vskip 2.5in
\caption{\xte\ spectrum of pulsed flux from \psr\ obtained by
subtracting the off-peak spectrum from the on-peak spectrum,
and fitting to an absorbed power-law model (see the text).}
\label{spectrum}
\end{figure}
\section{Discussion and Conclusions}

The discovery of \psr\ using \xte\
verifies the conjecture of \citet{ter08}
that \xmmu, \igr, and \tev\ are all manifestations of a
young pulsar/PWN system.  Even though the \xte\ field of view
is wider than those of \xmm\ and \chandra,
there is little doubt that \psr\ is the
compact source in \igr, given the morphological and spectral evidence,
and the compatibility of the \xte\ measured pulsed flux
with that of \xmmu. \citet{ter08} estimated that the
spin-down luminosity of \psr\ would be $\approx 9 \times 10^{36}$~erg~s$^{-1}$
based on the empirical correlation between $\dot E$ and
$\Gamma_{\rm PSR}$ of \citet{got03}; this turns out to have
been an accurate prediction.  

Assuming a distance of 7~kpc as proposed in Section 2.1, 
the X-ray luminosities of the pulsar and PWN,
$L_{\rm PSR}(2-10\,{\rm keV}) = 2.9 \times 10^{-3}\,\dot E\,d_{7}^2$
and $L_{\rm PWN}(2-10\,{\rm keV}) = 6.6 \times 10^{-4}\,\dot E\,d_{7}^2$,
respectively, are consistent with the range of
pulsar/PWN systems \citep{kar08}.
The $20-100$~keV flux of \igr\ measured
by \integral\ \citep[$2 \times 10^{-11}$
erg~cm$^{-2}$~s$^{-1}$;][]{ter08} corresponds to
$L(20-100\,{\rm keV}) = 0.012\,\dot E\,d_{7}^2$.
Given the $2-20$~keV flux and spectral information from \xmm\ and \xte,
the \integral\ source is likely to be dominated by the pulsar
rather than the PWN.

The TeV flux from \tev\
\citep[$2.2 \times 10^{-12}$ erg~cm$^{-2}$~s$^{-1}$;][]{ter08}
corresponds to $L(0.35-10\,{\rm TeV}) = 1.3 \times 10^{-3}\,\dot E\,d_{7}^2$.
Such a small efficiency of converting spin-down power to high-energy
radiation is typical of high $\dot E$ pulsars.
The ratio $F(0.35-10\,{\rm TeV})/F_{\rm PWN}(2-10\,{\rm keV}) \approx 2$
falls squarely on the inverse correlation between this quantity and $\dot E$
that was fitted by \citet{mat09b}, and modeled by them in terms of an
evolving PWN emitting synchrotron X-rays and inverse Compton 
$\gamma$-rays.  \psr\ is in transition from
a synchrotron dominated X-ray PWN to an inverse Compton scattered TeV nebula,
an expected phase through which a middle-aged pulsar will pass.

\acknowledgements

We thank the \xte\ project for making the time available for this
program, and the mission planners for carefully scheduling the observations.
This investigation is also based on observations obtained with \xmm,
an ESA science mission with instruments and contributions directly
funded by ESA Member States, and NASA.

\end{document}